\documentclass[reprint,aps,prl]{revtex4-1}
\usepackage{subfig}
\usepackage{graphicx}
\usepackage{amsmath}
\usepackage{units}
\usepackage[version=3]{mhchem}

\begin{document}

\title{Toward Quantitative Phase-field Modeling of Dendritic Electrodeposition}
\author{Daniel A. Cogswell}
\affiliation{Samsung Advanced Institute of Technology America, Cambridge, MA 02142, USA}
\date{\today}

\begin{abstract}
A thin-interface phase-field model of electrochemical interfaces is developed based on Marcus kinetics for concentrated solutions, and used to simulate dendrite growth during electrodeposition of metals. The model is derived in the grand electrochemical potential to permit the interface to be widened to reach experimental length and time scales, and electroneutrality is formulated to eliminate the Debye length. Quantitative agreement is achieved with zinc Faradaic reaction kinetics, fractal growth dimension, tip velocity, and radius of curvature. Reducing the exchange current density is found to suppress the growth of dendrites, and screening electrolytes by their exchange currents is suggested as a strategy for controlling dendrite growth in batteries.
\end{abstract}

\maketitle

Understanding the cause of dendrite growth during electrodeposition is a challenging problem with important technological relevance for advanced battery technologies  \cite{Park2014}. Controlling the growth of dendrites would solve a decades-old problem and enable the use of metallic electrodes such as lithium or zinc in rechargeable batteries, leading to significant increases in energy density.

Due to the complexity of observed deposition patterns \cite{Grier1986,Sawada1986,Kahanda1989,Trigueros1991}, a complete theoretical understanding of the formation of dendrites from binary electrolytes has not been developed. Modeling of electrodeposition has largely focused on analysis of diffusion equations without consideration of morphology \cite{Chazalviel1990,Elezgaray1998,Rosso2006,Monroe2003,Akolkar2013}, or variations of diffusion limited aggregation \cite{Mayers2012,Park2014} which are applicable only at the limit of very small currents, and which do not account for surface energy.

In contrast, the phase-field method \cite{Boettinger2002,Steinbach2009} has succeeded at quantitatively modeling dendritic solidification at the limit of zero reaction kinetics \cite{Karma1998,Karma2001, Echebarria2004}, but has had only limited application to electrochemical systems with Faradaic reactions at the interface. The advantage of the phase-field method is that boundaries are tracked implicitly, and interfacial energy, interface kinetics, and curvature-driven phase boundary motion are incorporated rigorously.

Phase-field models of electrochemical interfaces have recently been developed \cite{Guyer2004,Guyer2004a,Pongsaksawad2007,Shibuta2007,Okajima2010,Liang2012,Ely2014} and applied to dendritic electrodeposition \cite{Shibuta2007,Okajima2010,Ely2014}, but these models suffer from significant limitations. Perhaps the most serious oversight in current electrodeposition models is the assumption of linearized or Butler-Volmer kinetics. It has been known for several decades that even seemingly simple metal reduction reactions are in fact multi-step and limited by \textit{electron} transfer \cite{Mattsson1959,Epelboin1975}. As a consequence, curved Tafel plots that deviate from Butler-Volmer have been reported for zinc reduction \cite{Matsuda1979,Fawcett1990}.

Simulating experimental length and time scales is a second challenge. Guyer \textit{et al.} \cite{Guyer2004,Guyer2004a} provided a diffuse-interface description of charge separation at an electrochemical interface capable of modeling double layers and Butler-Volmer kinetics, but the model is essentially too complex for practical use. The evolution equations are numerically unstable and require high temporal and spatial resolution, limiting simulations to 1D. Shibuta \textit{et al.} \cite{Shibuta2007} addressed the length and time scale challenge with a thin-interface electrodeposition model, but did not implement Butler-Volmer reaction kinetics or apply the correct electroneutrality condition. These shortcomings were addressed in a follow-up paper \cite{Okajima2010}, although Butler-Volmer kinetics is merely approximated with nonlinear diffusivity.

This paper presents a phase-field model for electrodeposition that addresses both the reaction kinetics and the length and time scale issues. A consistent form of Marcus kinetics for concentrated solutions is incorporated, and the model is derived in the grand canonical ensemble \cite{Plapp2011} with an antitrapping current included \cite{Karma2001,Echebarria2004} to permit simulation of experimental length and time scales.

\textit{Free Energy Formulation} --
To show the relation to previous electrodeposition models, the phase-field model is presented first in terms of free energy, and then extended to the grand free energy so the interface can be widened for computational efficiency without introducing non-physical jumps in chemical potential \cite{Plapp2011}. The free energy functional for an electrochemical interface is \cite{Guyer2004,Guyer2004a,Garcia2004}:
\begin{equation}
 F\left[\xi,c_i,\phi\right]=\int_V\left[f(\xi,c_i)+\frac{1}{2}\kappa(\vec\nabla\xi)^2+\rho\phi\right]\, dV
 \label{Eq:energy_functional}
\end{equation}
where $\xi$ is an order parameter that distinguishes the electrode ($\xi=1$) from the electrolyte ($\xi=0$), $c_i$ are the mole fractions of the chemical species (for a binary system, anions, cations, and a neutral species), $\phi$ is the electric potential, $f(\xi,c_i)$ is the homogeneous Helmholtz free energy density, $\kappa$ is the gradient energy coefficient, and $\rho=\sum_iz_i\mathcal{F}c_i$ is charge density.

The homogeneous free energy $f(\xi,c_i)$  is an interpolation between the free energies of the electrode and electrolyte, which are assumed here to be ideal solutions:
\begin{equation}
\begin{split}
 f(\xi,c_i)&=f^s p(\xi)+f^l(1-p(\xi))+Wg(\xi) \\
 f^{s,l}&=\sum_{i=1}^Nc_i\mu_i^{\circ s,l}+RTc_i\ln(c_i)
\end{split}
 \label{Eq:free_energy}
\end{equation}
$\mu_i^{\circ s,l}=-RT\ln(c_i^{\circ s,l})$ are the chemical potentials of the pure components, $p(\xi)=\xi^3(6\xi^2-15\xi+10)$ is an interpolation function,  $g(\xi)=\xi^2(1-\xi)^2$ is a double-well function, and $W$ sets the height of the energy barrier between the phases. The physical quantities of surface energy $\gamma$ and interfacial width $\delta$ are related to the model parameters and choice of $g(\xi)$ \cite{Cahn1959} according to $\gamma=\sqrt{\frac{\kappa W}{18}}$ and $\delta=\sqrt{\frac{8\kappa}{W}}$.

\textit{Grand Canonical Formulation} --
A problem now arises if $\delta$ is chosen to be larger than the physical width of the interface, which may only be a few nanometers. If the interfacial points interpolate between two free energies at the same composition, as in Eq. \ref{Eq:free_energy}, the energy of the interfacial points lie above the common tangent line. As a result, widening the interface for computational necessity adds more non-equilibrium material, creating a non-physical jump in chemical potential and exaggerated solute trapping. This issue has been a recent focus of phase-field modeling, leading to so-called thin interface formulations that eliminate these non-physical effects \cite{Kim1999,Karma2001,Echebarria2004,Folch2005}.

Plapp recently showed that thin-interface formulations can be unified with a model derived in the grand canonical ensemble \cite{Plapp2011}. Following his approach, the grand free energy functional for an electrochemical system is:
\begin{equation}
 \begin{split}
  \Omega\left[\xi,\hat{\mu}_i,\phi\right]=\int &w^sp(\xi)+w^l(1-p(\xi)) \\
  &+Wg(\xi)+\frac12\kappa(\nabla\xi)^2+\rho\phi\, dV
 \end{split}
\end{equation}
where $w^s$ and $w^l$ are the homogeneous grand energy densities of the solid electrode and liquid electrolyte, respectively. Compared with the free energy functional, the grand energy functional exchanges concentration $c_i$ for chemical potential $\mu_i$ as the natural variable. As Plapp noted, equilibrium between phases involves the intensive variable chemical potential, but equations of motion are derived for concentration, the conjugate variable. As a result, alloy phase-field models formulated in terms of a phase variable and concentration do not necessarily establish constant chemical potential at equilibrium.

Treating $\hat\mu$ as the natural variable has an additional numerical benefit for simulation at low electrolyte concentrations. As $c_i\rightarrow 0$, the slope of the free energy curves becomes steep due to entropy, and very small fluctuations in $c_i$ lead to large changes in energy, causing numerical instability. This phenomenon appears to have restricted the range of feasible electrolyte compositions in other phase-field models \cite{Shibuta2007,Okajima2010}.  With $\hat\mu$ as the natural variable however, energy changes are much less sensitive to fluctuations, and much more robust at low $c_i$.

The grand energies are found from a Legendre transform of the free energies, $w=f-\sum_{i=1}^Nn_i\mu_i=f-\sum_{i=1}^Nc_i\hat{\mu}_i$, where $n_i$ is the number of moles of component $i$ and $\mu_i=\frac{\partial f}{\partial n_i}$ is its chemical potential. For a system with a fixed number of substitutional atomic sites, $c_i$ is the mole fraction of component $i$, and $\hat{\mu}_i=\frac{\partial f}{\partial c_i}$ is its diffusion potential \footnote{This is a subtle but important difference from the diffusion equation of Guyer \textit{et al.} \cite{Guyer2004,Guyer2004a}, who treated the diffusion potential as the chemical potential.}, a difference in chemical potentials \cite{Cogswell2011}.  For an ideal solution, the homogeneous grand free energies are thus $w^{s,l}=\mu_N^{\circ s,l}+RT\ln(c_N^{s,l})$, where $N$ is the neutral component defined by a mole fraction constraint.

Thermodynamic equilibrium between two phases implies that the diffusion potential of each component is the same in both phases: $\hat{\mu}_i=\frac{\partial f^s}{\partial c_i}=\frac{\partial f^l}{\partial c_i}$. The diffusion potentials for an ideal solution (Eq. \ref{Eq:free_energy}) are $\hat{\mu}_i=\mu_i^{\circ s,l}-\mu_N^{\circ s,l}+RT\ln\left(\frac{c_i^{s,l}}{c_N^{s,l}}\right)$, which can be inverted to obtain the equilibrium concentration in each phase:
\begin{equation}
 c_i^{s,l}(\hat\mu)=\frac{e^{(\hat{\mu}_i-\epsilon_i^{s,l})/RT}}{1+\sum_{j=1}^{1-N}e^{(\hat{\mu}_j-\epsilon_j^{s,l})/RT}}
\end{equation}
where $\epsilon_i^{s,l}=\mu_i^{\circ s,l}-\mu_N^{\circ s,l}$. The total concentration is an interpolation between the two equilibrium concentrations: $c_i=(1-p(\xi))c_i^l+p(\xi)c_i^s$.

\textit{Reaction Kinetics} --
When a voltage is applied across the interface, Faradaic reactions occur and a current is generated. Reaction kinetics are incorporated into the phase evolution equation by matching the velocity of the sharp-interface limit of the phase-field model to the current-overpotential equation:
\begin{equation}
 i=i_0\left(e^{-\alpha n\mathcal{F}\eta/RT}-e^{(1-\alpha)n\mathcal{F}\eta/RT}\right)
 \label{Eq:current-overpotential}
\end{equation}
where $i_0$ is the exchange current density, $\eta$ is overpotential, and $\alpha$ is the transfer coefficient,  defined according to the Marcus theory of electron transfer \cite{Henstridge2012,Bazant2013} as $\alpha=\frac{1}{2}+\frac{n\mathcal{F}\eta}{2\lambda}$, where $\lambda$ is the reorganization energy. Marcus kinetics, which has been measured for zinc \cite{Matsuda1979}, is an approximation at small overpotentials of Marcus-Hush-Chidsey kinetics \cite{Zeng2014}.  The exchange current density is assumed to be constant, a reasonable assumption for metals such as zinc where the exchange current is insensitive to electrolyte concentration \cite{Guerra2004}, a consequence of a rate-limiting step which does not involve a solvated ion \cite{Epelboin1975}.

Overpotential is defined variationally as a local field quantity following other phase-field models of electrokinetics \cite{Cogswell2012, Bazant2013}:
\begin{equation}
 \frac{n\mathcal{F}}{V_m}\eta[\xi,\hat\mu_i]=\frac{\delta\Omega}{\delta \xi}=Wg'(\xi)+p'(\xi)(\Delta\omega+\phi\Delta \rho)-\vec\nabla\cdot\kappa\vec\nabla\xi
 \label{Eq:overpotential}
\end{equation}
where $\Delta\omega=\omega^s-\omega^l$ and $\Delta\rho=\rho^s-\rho^l$. The total interfacial overpotential is an integral of this field across an interface, $\eta_t=\frac{1}{A\delta}\int\eta[\xi,\hat\mu_i]\,dV$, where $A$ is the area of the interface and $A\delta$ is the volume of the diffuse interface.

\begin{figure}[t]
 \includegraphics[width=.85\columnwidth]{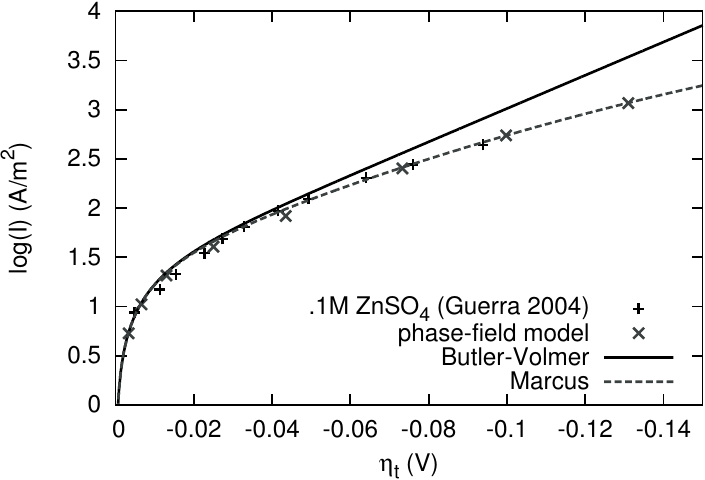}
 \caption{Tafel plot for zinc deposition from a \unit[.1]{M} \ce{ZnSO4} solution showing agreement between the phase-field model with an interfacial width of $\unit[1]{\mu m}$, the sharp interface limit (Eq. \ref{Eq:current-overpotential}, Marcus kinetics), and experimental observation \cite{Guerra2004}.}
 \label{Fig:Tafel}
\end{figure}

The phase-field evolution equation is then found by matching the velocity of the sharp interface limit of the phase equation to Eq. \ref{Eq:current-overpotential} \cite{Liang2012,Ely2014}. The evolution equation is:
\begin{equation}
 \frac{\partial\xi}{\partial t}=\frac{V_m\gamma}{n\mathcal{F}\kappa}i_0\left(e^{-\alpha n\mathcal{F}\eta[\xi,\hat\mu_i]/RT}-e^{(1-\alpha)n\mathcal{F}\eta[\xi,\hat\mu_i]/RT}\right)
 \label{Eq:phase_equation}
\end{equation}
Fig. \ref{Fig:Tafel} illustrates that this kinetic equation for the diffuse interface accurately reproduces the Tafel behavior of Eq. \ref{Eq:current-overpotential}.

\textit{Diffusion} --
Evolution equations for $\hat\mu_i$ are derived from the conservation law $\frac{\partial c_i}{\partial t}=-\vec\nabla\cdot\vec J_i$ by recognizing that $c_i$ is a function of $\xi$ and $\hat\mu_i$ in the grand ensemble: 
\begin{equation}
 \frac{\partial c_i}{\partial t}=-\vec\nabla\cdot\vec{J}_i=\frac{\partial c_i}{\partial\xi}\frac{\partial\xi}{\partial t}+\frac{\partial c_i}{\partial\hat{\mu}_i}\frac{\partial\hat{\mu}_i}{\partial t}
\end{equation}
This equation can be rearranged to express the time evolution of $\hat{\mu}_i$ as:
\begin{equation}
 \frac{\partial\hat{\mu}_i}{\partial t}=-\frac{1}{\chi_i}\left(\vec\nabla\cdot\vec{J}_i+p'(\xi)\left[c_i^s-c_i^l\right]\frac{\partial\xi}{\partial t}\right)
 \label{Eq:mu_evolution}
\end{equation}
with $\chi_i=\frac{\partial c_i}{\partial\hat{\mu}_i}$ and $\vec{J}_i=-M_i\left[c_N\vec\nabla\hat\mu_i+z_i\mathcal{F}\vec\nabla\phi\right]+\vec{J}_{at}+\vec{q}$, where where $M_i=\frac{D_ic_i}{RT}$ according to the Nernst-Einstein relation, $\vec{J}_{at}=-\sqrt{\frac{\kappa}{2W}}\left(c_i^l-c_i^s\right)\frac{\partial \xi}{\partial t}\frac{\vec\nabla\xi}{\left|\vec\nabla\xi\right|}$ is an antitrapping current that eliminates excessive solute trapping at the interface \cite{Karma2001,Echebarria2004}, and $\vec q$ is a Langevin noise term accounting for thermal fluctuations \cite{Karma1998}. A derivation of this flux equation is presented in Supplemental Material.

\begin{table}[t]
 \begin{tabular}{cccccc}
  \hline
  Variable & Description & Value & Source  \\
  \hline
  $n$ & electrons transferred & 2 & \cite{Epelboin1975} \\
  $\gamma$ & surface energy energy & $\unit[.5]{J/m^2}$ & \cite{Bilello1983}  \\
  $V_m$ & molar volume & $\unit[9.16]{cm^3}$ & \cite{Singman1984} \\
  $D$ & mutual diffusivity & $\unit[3.68\times 10^{-10}]{m^2/s}$ & \cite{Albright1975} \\
  $t_+$ & transference number & .4 & \cite{Dye1960} \\
  $i_0$ & exchange current density & $\unit[28]{A/m^2}$ & \cite{Guerra2004} \\
  $\alpha$ & transfer coefficient & .5 & \cite{Guerra2004} \\
  $\lambda$ & reorganization energy & \unit[120]{kJ/mol} & \cite{Matsuda1979} \\
  \hline
 \end{tabular}
 \caption{Parameters employed to model zinc electrodeposition from a binary \ce{ZnSO4} electrolyte.}
 \label{Tbl:parameters}
\end{table}

\textit{Electroneutrality} --
At this point the interface can now be widened without introducing a jump in chemical potential. However, Poisson's equation still places a severe practical restriction on the width of the interface, since the Debye length is typically on the order of \unit[1]{nm}. Thus is it necessary to ignore effects of the double-layer structure and to assume electroneutrality. Experimental observations support the assumption that it is not necessary to consider space-charge effects when considering the stability of electrodeposits \cite{Elezgaray1998}. An additional benefit of electroneutrality is the simplification of the model so that it is only necessary to explicitly track the movement of cations.

Importantly, electroneutrality does not imply that Laplace's equation holds in place of Poisson's equation. Instead, $\phi$ must be found from an expression for current conservation, $\frac{n\mathcal{F}}{V_m}\left(\frac{\partial c_+}{\partial t}-\frac{\partial c_-}{\partial t}\right)=-\vec\nabla\cdot i=0$, with the following constraints introduced by electroneutrality $c=c_+=c_-$, $n=z_+=-z_-$, and $\hat\mu=\hat{\mu}_+=\hat{\mu}_-$. The electroneutrality condition becomes:
\begin{equation}
 \vec\nabla\cdot\bigg((D_+-D_-)c\, c_N\vec\nabla\hat{\mu}\bigg)+\vec\nabla\cdot\bigg(n\mathcal F(D_++D_-)c\vec\nabla\phi\bigg)=0
 \label{Eq:electroneutrality}
\end{equation}
where $D_+$ and $D_-$ are the diffusivities of the cations and anions in the electrolyte \footnote{$D_+$ and $D_-$ can be obtained from transference number $ t_+=\frac{z_+D_+}{z_+D_+-z_-D_-}$ and the ambipolar diffusivity $D=\frac{D_+D_-(z_+-z_-)}{z_+D_+-z_-D_-}$.}. Additionally, the application of electroneutrality to Eq. \ref{Eq:overpotential} implies that $\rho^l=0$ and $\rho^s=-n\mathcal{F}c^s$, so that $\Delta\rho$ represents the electrons required to create neutral $c^s$ from $c_+^l$ ions in the electrolyte.

\begin{figure}[t]
 \subfloat[Da=.1]{\includegraphics[width=.3\columnwidth]{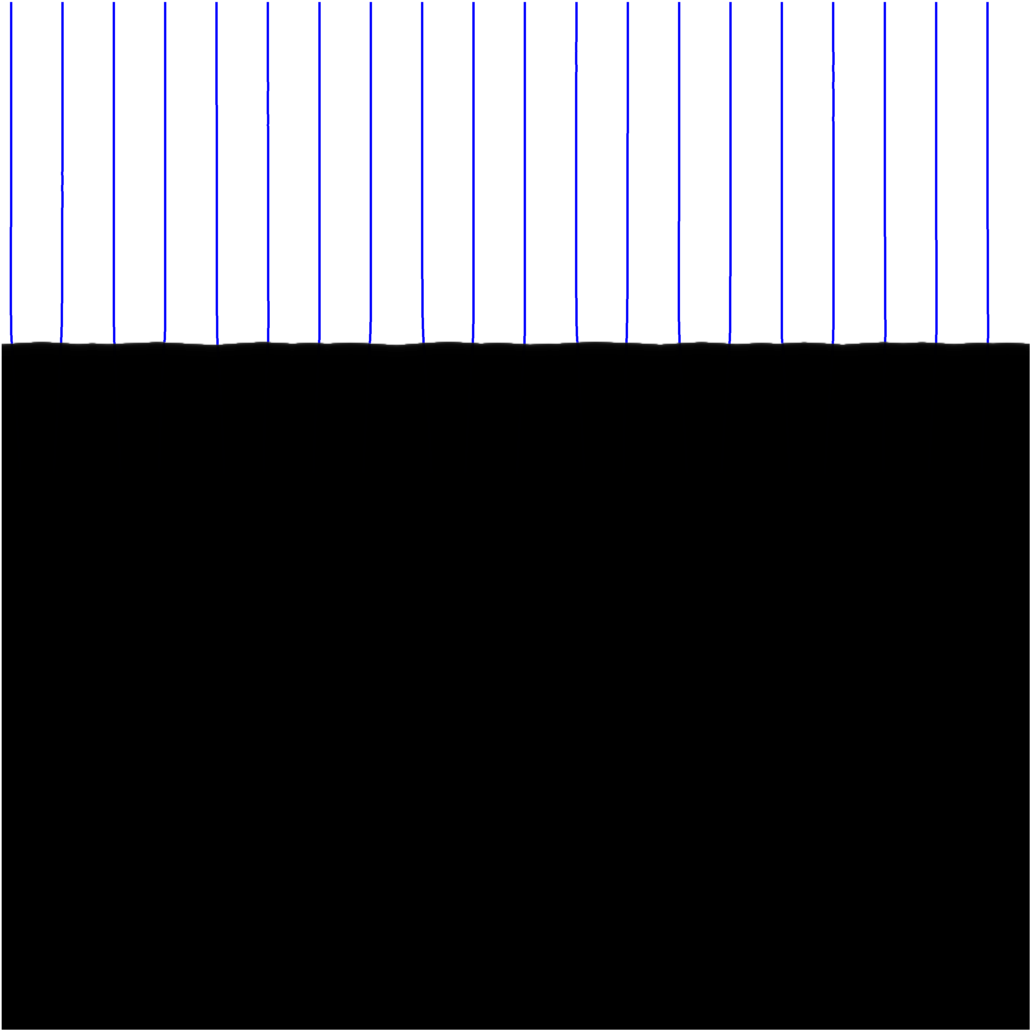}}\hspace{.02\columnwidth}
 \subfloat[Da=1]{\includegraphics[width=.3\columnwidth]{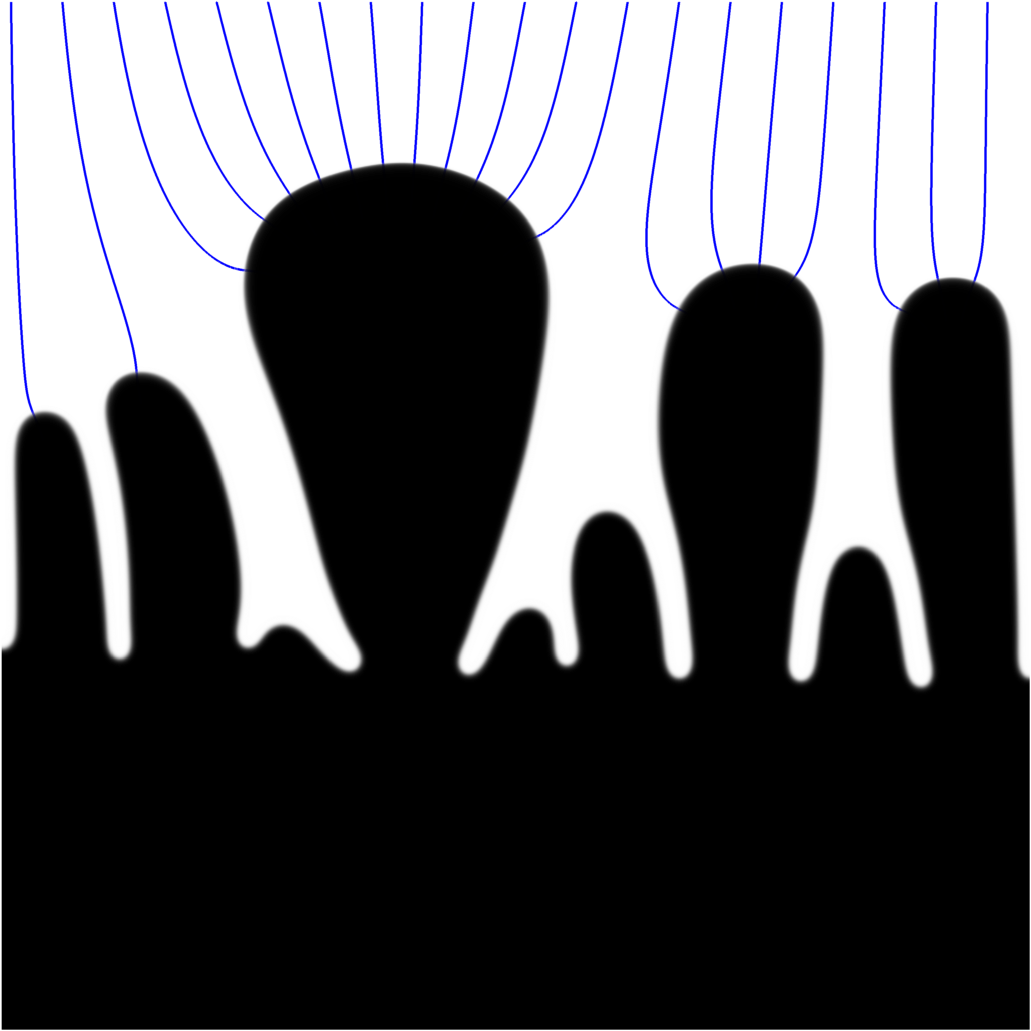}}\hspace{.02\columnwidth}
 \subfloat[Da=10]{\includegraphics[width=.3\columnwidth]{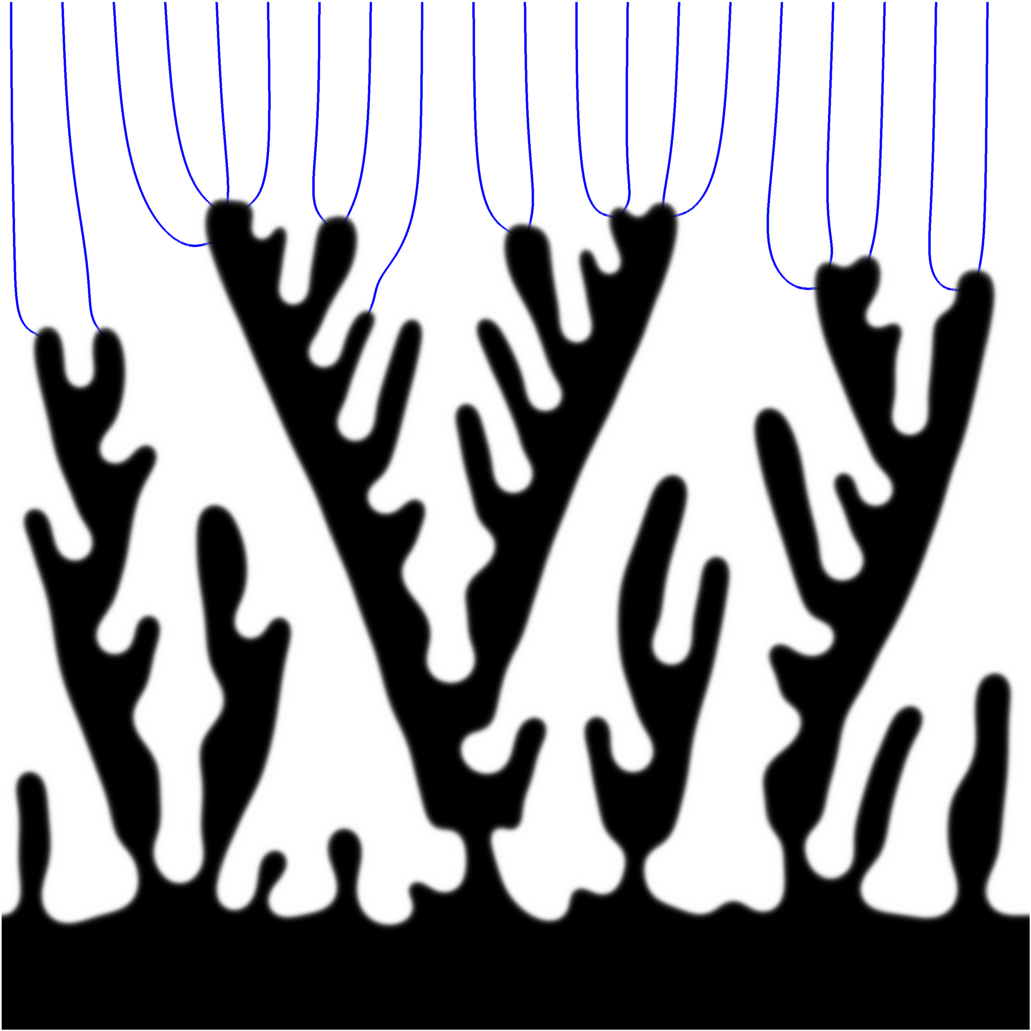}}\hspace{.02\columnwidth}
\caption{Simulated electrodeposition morphologies from a from a 1M \ce{ZnSO4} electrolyte at different Damkohler numbers. The width of each simulation is $\unit[150]{\mu m}$, and electric field lines are shown in blue.}
\label{Fig:dendrites}
\end{figure}

\textit{Computation} --
The model was made non-variational by changing the interpolating function in Eq. \ref{Eq:mu_evolution} to $p(\xi)=\xi$ for numerical efficiency, and as required for the antitrapping current \cite{Karma1998,Karma2001,Echebarria2004,Plapp2011}. Because zinc has a hexagonal crystal structure that strongly affects dendrite morphology \cite{Grier1986,Sawada1986}, six-fold anisotropy in the interfacial energy was implemented using the standard approach \cite{Kobayashi1993} with  $\gamma(\theta)=\gamma\left[1+\epsilon_6\cos(6\theta)\right]$, where $\theta$ is the angle between the surface normal and the crystallographic axes, and $\epsilon_6=.01$ sets the strength of the anisotropy. The evolution equations (Eq. \ref{Eq:phase_equation}, \ref{Eq:mu_evolution}, and \ref{Eq:electroneutrality}) were solved using multigrid techniques detailed in Supplemental Material. Other simulation parameters are presented in Table \ref{Tbl:parameters}.

The model was parameterized in terms of a dimensionless Damkohler number, which expresses the relative importance of the reaction rate to diffusion, $Da=\frac{i_0V_m/n\mathcal{F}}{D/L}$, where $i_0$ is the exchange current density, $n$ the number of electrons transferred, $D$ the electrolyte diffusivity, and $L$ the distance between the two electrodes. 2D simulations were performed for direct comparison with experimental morphologies obtained from 2D thin-cell geometries.

\textit{Results} --
Fig. \ref{Fig:Tafel} shows the success of the phase-field model at reproducing Marcus kinetics while addressing the length scale challenge. The interface was widened by roughly three orders of magnitude to $\unit[1]{\mu m}$, yet the underlying nonlinear kinetics occurring at the scale of the electric double layer were accurately reproduced.

Fig. \ref{Fig:dendrites} examines the effect of the Damkohler number on dendrite growth morphology, revealing that low Damkohler numbers have a dramatic effect on suppressing the formation of dendrites. Reaching a kinetically limited regime before reaching a transport-limited regime is like imposing a speed limit on the velocity of the interface, lessening the disparate interface velocities that lead to dendrites. Dendrites grow when the electric field concentrates at protrusions, increasing the local overpotential and enhancing growth. As dendrites grow taller they attract more electric field lines and screen their shorter neighbors, whose growth eventually ceases (see video in Supplemental Material).

Surface energy anisotropy plays an important role in growth morphology as well \cite{Grier1986}. Zinc has a hexagonal crystal structure and tends to grow branching or fractal dendrites, while lithium, with a cubic crystal structure (less inherent anisotropy), grows needle-like dendrites. Simulation with 4-fold anisotropy indeed produces needles (see Supplemental Material).

In the diffusion limited regime, fractal dimension has proved to be a reliable measure for electrodeposits, with that of zinc consistently measured in the range 1.60-1.75 \cite{Grier1986,Argoul1988,Kahanda1989,Trigueros1991}. Using the box counting method \cite{Karperien1999-2013}, the fractal dimension of Fig. \ref{Fig:dendrites} (c) was found to be 1.67, showing the capability of the phase-field model to capture fractal growth phenomena.

\begin{figure}[t]
 \includegraphics[width=.9\columnwidth]{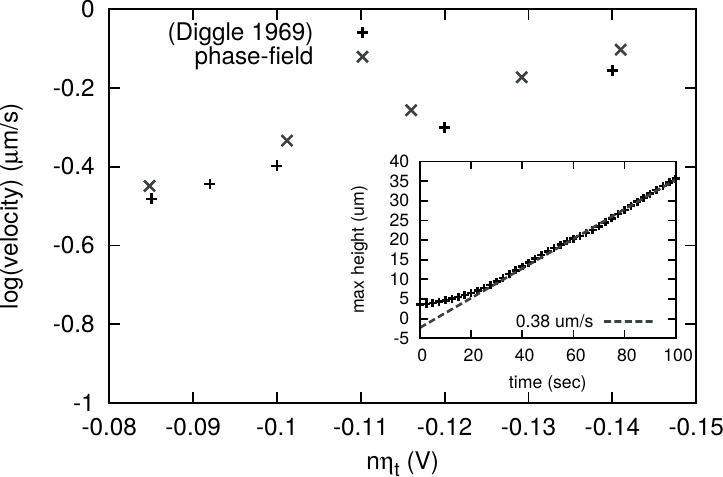}
\caption{Comparison of simulated and experimental dendrite tip velocity as a function of overpotential for deposition of zinc from a $\unit[.1]{M}$ zincate electrolyte. An exchange current of $i_0=\unit[1000]{A/m^2}$ was used for zincate, following \cite{Diggle1969}, and $Da=10$. The inset figure shows the hight of a simulated dendrite tip ($n\eta_t=\unit[84]{mV}$) with time.}
\label{Fig:velocity}
\end{figure}

After an initial formation stage, zinc dendrite tips are observed to grow at a constant velocity that depends exponentially on the applied overpotential \cite{Diggle1969}. Fig. \ref{Fig:velocity} shows agreement between experimental tip velocity measurements and  phase-field simulations of single dendrites grown from a perturbation. The inset figure shows the height of a simulated dendrite vs time, revealing that the dendrite grew at constant velocity. Since the velocity $v$ is proportional to the tip current, a linear relationship between $\log(v)$ and $\eta_t$ exists, as shown in Fig. \ref{Fig:velocity}, with the expected Tafel slope of $\alpha=.5$. 

Zinc dendrite tips are also know to be a parabolic with a characteristic radius of curvature \cite{Diggle1969}.  The simulations in Fig. \ref{Fig:velocity} produced parabolic tips with curvatures ranging from $\unit[.85]{\mu m}$ to $\unit[1.15]{\mu m}$, within the rage measured by Diggle \textit{et al.} \cite{Diggle1969} (see Table VIII). Details of how the curvature was measured are in Supplemental Material. Importantly, tip curvature cannot be predicted from models such as DLA that do not account for surface energy.

\textit{Discussion} --
Preventing dendrite growth by improving electrolyte transport in batteries (the denominator of the Damkohler number) has been demonstrated recently \cite{Park2014}, but little effort has been spent targeting the exchange current, despite the fact that kinetics are known to vary by orders of magnitude with slight changes in electrolyte composition \cite{Bard1973}. Surprisingly, reducing the exchange current to smoothen deposits appears to have been reported in a different context for cadmium decades ago \cite{Sotnikova1950}, and may also explain why magnesium, with an exchange current orders of magnitude smaller than lithium or zinc \cite{Peled1977}, is not observed to grow dendrites \cite{Matsui2011}. Recently it was observed that a small amount of bismuth at zinc surface inhibits dendrite growth \cite{Gallaway2014}, which might also be related to reaction kinetics.

Finally, there appear to be many similarities between electrodeposition and the phenomenon of viscous fingering \cite{Homsy1987}. In addition to visually similar morphologies, the growth process of both occur via mechanisms of shielding, spreading, and tip splitting. The exchange current in electrodeposition appears to act as a stabilizing force in an analogous way to gravitational stabilization of viscous fingering.

In conclusion, a phase-field model of electrochemical interfaces was developed to study the growth of dendrites during electrodeposition. The model was derived in the grand canonical ensemble to allow the interface to be widened to simulate experimental length and time scales, and Faradaic reactions were modeled rigorously with Marcus kinetics. Damkohler number, overpotential, and electrolyte concentration were investigated, and the model accurately reproduced the reaction kinetics, fractal dimension, and tip velocity and curvature of zinc dendrites. The results suggest that engineering the electrolyte to decrease the reaction kinetics could be a successful strategy for controlling dendrite growth.

\begin{acknowledgments}
I would like to thank M. Plapp for kindly discussing his grand potential phase-field formulation,  R.B. Smith, E. Khoo, and M.Z. Bazant for their critical reading of the text, and N.M. Schneider for helpful discussions of dendrite formation.
\end{acknowledgments}

\bibliography{electrodeposition}
\end{document}

\title{Supplemental Material: Toward Quantitative Phase-field Modeling of Dendritic Electrodeposition}
\author{Daniel A. Cogswell}
\affiliation{Samsung Advanced Institute of Technology America, Cambridge, MA 02142, USA}
\date{\today}

\maketitle

\section{Flux equations}
The free energy functional for electrochemical systems which was presented in the main text is:
\begin{equation}
 F\left[\xi,c_i,\phi\right]=\int_V\left[f(\xi,c_i)+\frac{1}{2}\kappa(\vec\nabla\xi)^2+\rho\phi\right]\, dV
 \label{Eq:energy_functional}
\end{equation}
As noted by Plapp \cite{Plapp2011} and highlighted in the main text, thermodynamic equilibrium is established by the intensive variable chemical potential, while conversation laws are applied to its conjugate variable, concentration. Thus although the model is formulated in terms of the diffusion potential, it is still necessary to determine an expression for the flux of concentration in terms of the diffusion potential $\hat\mu$.

The ternary diffusion equations for phase-field models have been derived in previous work to obey the Gibbs-Duhem relation \cite{Nauman2001,Cogswell2011}. These equations applied to a binary electrolyte are:
\begin{subequations}
 \begin{equation}
  \vec{J}_+=-\frac{D_+c_+}{RT}\bigg[(1-c_+)\vec\nabla\frac{\delta F}{\delta c_+}-c_-\vec\nabla\frac{\delta F}{\delta c_-}\bigg]
 \end{equation}
 \begin{equation}
  \vec{J}_-=-\frac{D_-c_-}{RT}\bigg[(1-c_-)\vec\nabla\frac{\delta F}{\delta c_-}-c_+\vec\nabla\frac{\delta F}{\delta c_+}\bigg]
 \end{equation}
\end{subequations}
The variational derivatives are $\frac{\delta F}{\delta c_i}=\hat\mu_i+z_i\mathcal{F}\phi$, where $\hat\mu_i$ is the diffusion potential. Upon substitution, the flux equations become:
\begin{subequations}
 \begin{align}
  \begin{split}
   \vec{J}_+=-\frac{D_+c_+}{RT}\bigg[&(1-c_+)\vec\nabla\hat\mu_+-c_-\vec\nabla\hat\mu_-\\
   &+(z_+-z_+c_+-z_-c_-)\mathcal{F}\vec\nabla\phi\bigg]
  \end{split}
 \end{align}
 \begin{align}
  \begin{split}
   \vec{J}_-=-\frac{D_-c_-}{RT}\bigg[&(1-c_-)\vec\nabla\hat\mu_--c_+\vec\nabla\hat\mu_+\\
   &+(z_--z_-c_--z_+c_+)\mathcal{F}\vec\nabla\phi\bigg]
  \end{split}
 \end{align}
 \label{Eq:ternary_flux}
\end{subequations}

\section{Electroneutrality}
For a binary electrolyte with an assumption of electroneutrality, it is only necessary to track one of the charged species. Electroneutrality implies the following relationship holds in the electrolyte:
\begin{equation}
 \frac{n\mathcal{F}}{V_m}\left(\frac{\partial c_+}{\partial t}-\frac{\partial c_-}{\partial t}\right)=-\vec\nabla\cdot i=0
 \label{Eq:electroneutrality}
\end{equation}
with the constraints that $c=c_+=c_-$, $n=z_+=-z_-$, and $\hat\mu=\hat{\mu}_+=\hat{\mu}_-$. Substitution of these constraints into the ternary flux equations (Eq. \ref{Eq:ternary_flux}) produces:
\begin{subequations}
 \begin{equation}
  \vec{J}_+=-\frac{D_+c}{RT}\bigg[c_N\vec\nabla\hat\mu+z_+\mathcal{F}\vec\nabla\phi\bigg]
 \end{equation}
 \begin{equation}
  \vec{J}_-=-\frac{D_-c}{RT}\bigg[c_N\vec\nabla\hat\mu+z_-\mathcal{F}\vec\nabla\phi\bigg]
 \end{equation}
\end{subequations}
where $c_N=1-2c$ is the mole fraction of the neutral component. Substitution of these fluxes into Eq. \ref{Eq:electroneutrality} results in and equation that can be solved numerically for $\phi$:
\begin{equation}
 \vec\nabla\cdot\bigg((D_+-D_-)c\, c_N\vec\nabla\hat\mu\bigg)+\vec\nabla\cdot\bigg(n\mathcal F(D_++D_-)c\vec\nabla\phi\bigg)=0
\end{equation}
Eq. 5 in the main text can then be used to convert from $\hat\mu$ to $c$ as necessary.

\section{Computation}
The evolution equations (Eq. 7, 9, and 10 in the main text) are solved for the variables $\xi$, $\hat\mu$, and $\phi$ using Crank-Nicolson time integration, a finite volumes discretization on a regular grid, and a full approximation scheme (FAS) multigrid method \cite{Trottenberg2001}. A Red-Black Gauss-Seidel smoother was used with F-cycles and 1 pre- and 2 post-smoothing iterations. The full-weighting transfer operator was used to restrict both the defect and the state variables. Smoothing of the state variables is necessary to avoid instability on coarse grids resulting from the strongly varying diffusivity between the solid and the liquid \cite{Brandt2011}.

Dirichlet boundaries were applied at the top and bottom of the simulation, and Neumann conditions were applied on the sides. A Neumann condition on $\phi$ at the top boundary was used to impose a constant electric field and generate a current.

To implement the Damkohler number computationally, the electrode separation distance and electrolyte diffusivity are held constant, and the exchange current is varied. A more physically realistic approach would hold the diffusivity constant and vary the electrolyte separation, but this significantly increases computational cost.

\section{Calculation of tip curvature}

\begin{figure}[t]
 \includegraphics[width=.9\columnwidth]{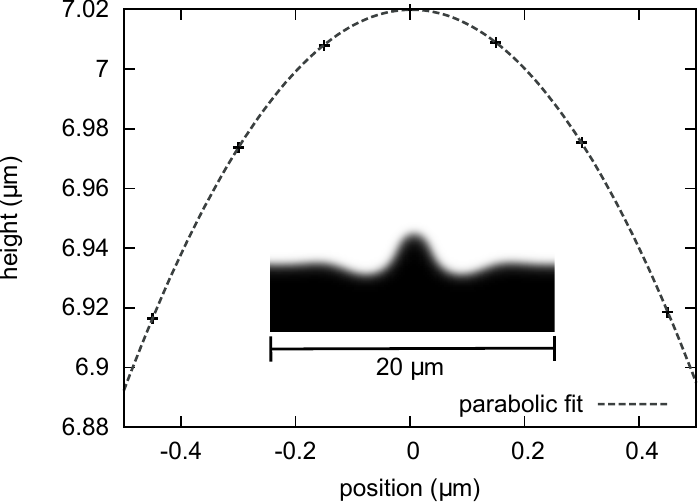}
\caption{Interfacial points in the vicinity of a growing dendrite are determined by interpolation. the dotted line is a second-order polynomial that has been fitted to these points. The inset image shows a wider view of the same dendrite tip. This dendrite was grown under an imposed overpotential of $n\eta_t=\unit[116]{mV}$, and has a radius of curvature at the tip of $r=\unit[.99]{\mu m}$.}
\label{Fig:curvature}
\end{figure}

Single dendrites were grown by placing a small perturbation in the middle of the simulation frame and curvature measurements were made as it grew. The interface is defined by $\xi=.5$, which was found by interpolating $\xi$ between grid points. As discussed elsewhere \cite{Karma1998}, this interpolation is important since the measured curvature is very sensitive to the interfacial position at the tip. Thus the interfacial points in the vicinity of the growing tip were then fit with a second-order polynomial, as shown in Fig. \ref{Fig:curvature}, and the radius of curvature was obtained from this polynomial according to the equation:
\begin{equation}
 k=\frac 1r=\frac{y''}{\left(1+y'^2\right)^\frac32}
\end{equation}
where $y(x)$ is the polynomial fitted to the dendrite tip, and the radius of curvature $r$ is calculated at $x=0$.

\section{Dendrite growth morphology}
\begin{figure*}[p]
 \subfloat[]{
 \begin{tabular}{ccccc}
  \includegraphics[width=.18\textwidth]{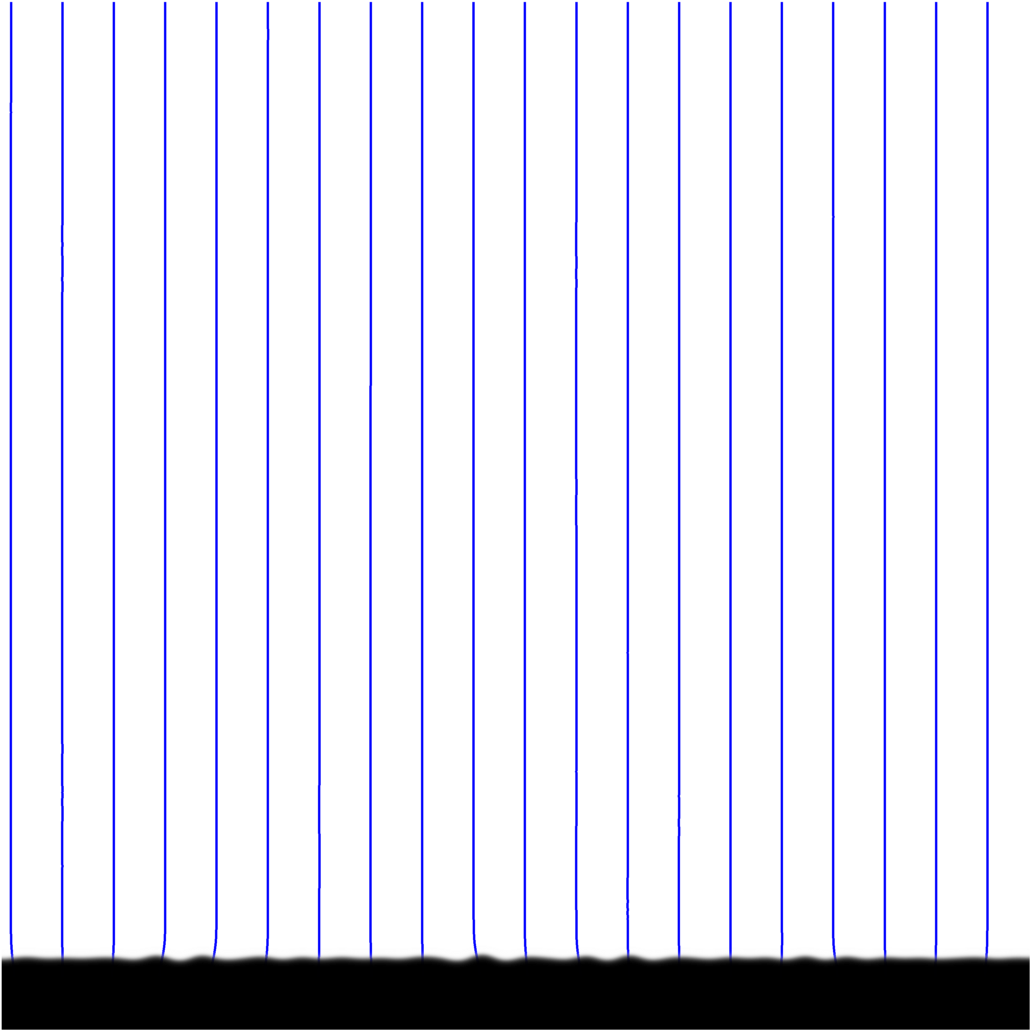} &\includegraphics[width=.18\textwidth]{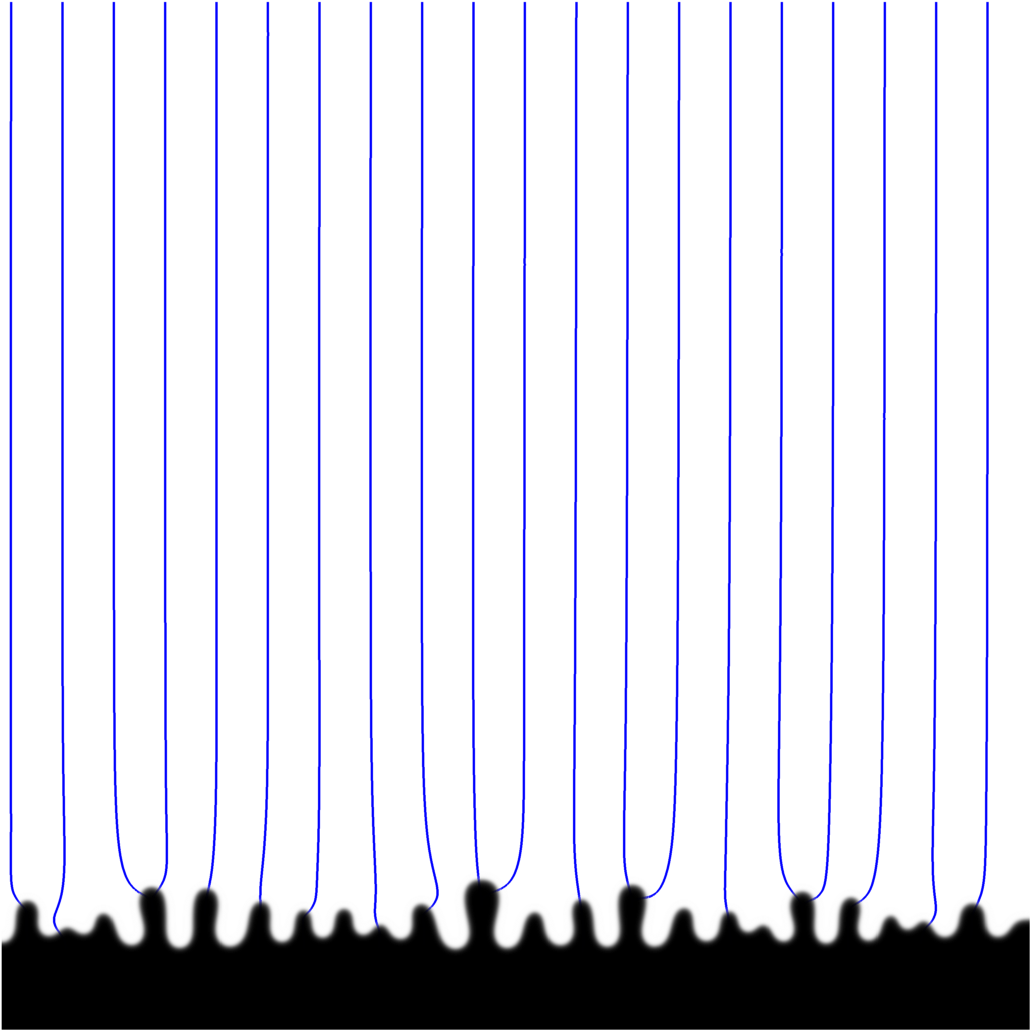} & \includegraphics[width=.18\textwidth]{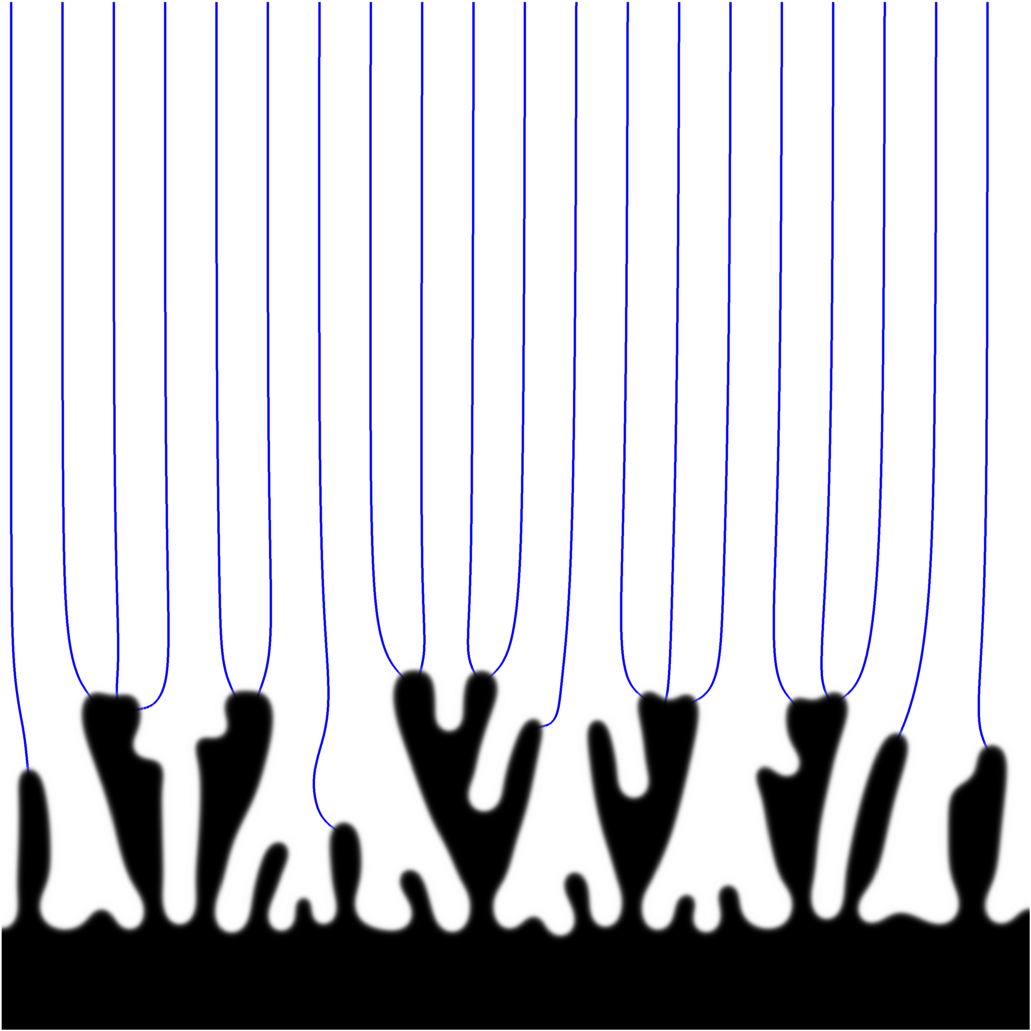} & \includegraphics[width=.18\textwidth]{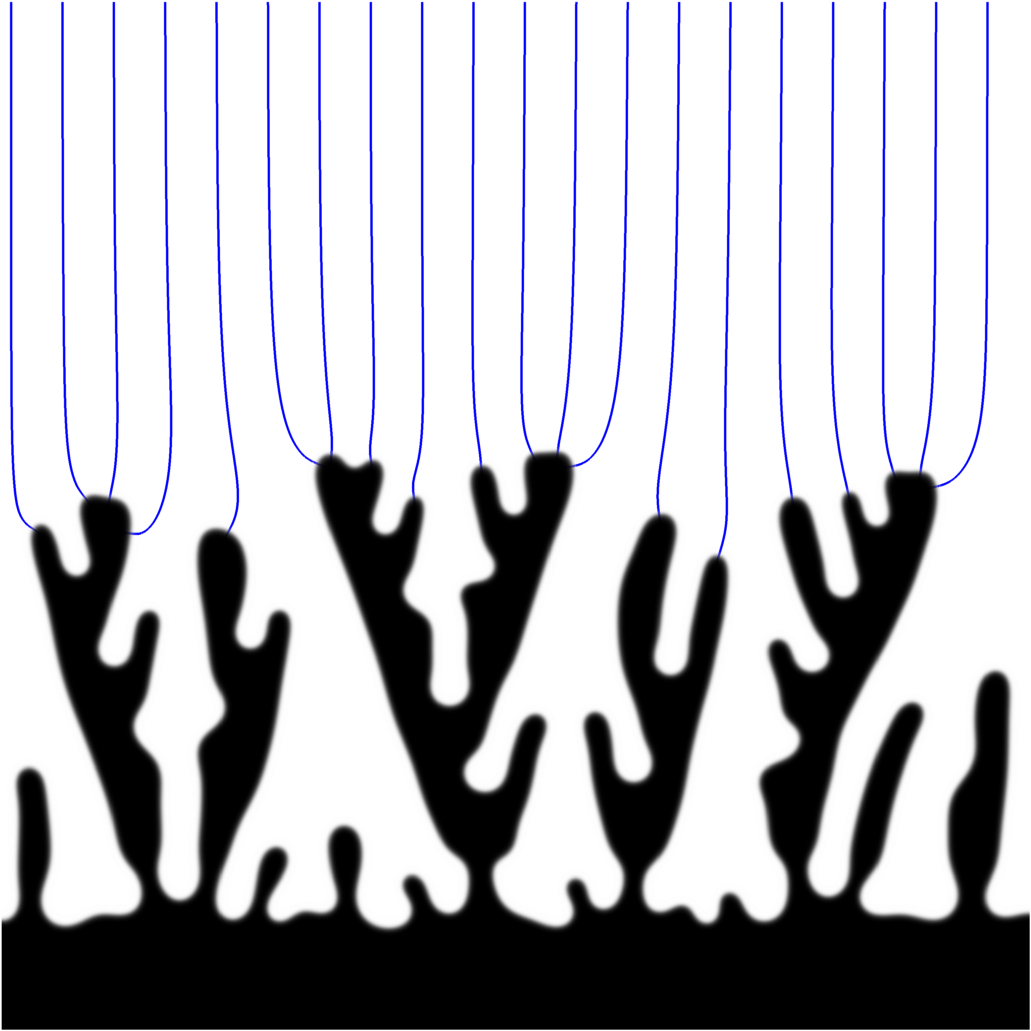} & \includegraphics[width=.18\textwidth]{E1000-D10-fl020000.png}  \\
 t=\unit[30]{s} & t=\unit[50]{s} & t=\unit[100]{s} & t=\unit[150]{s} & t=\unit[200]{s} \\
 \end{tabular}
\label{Fig:hexagonal}}\\
 \subfloat[]{
 \begin{tabular}{ccccc}
  \includegraphics[width=.18\textwidth]{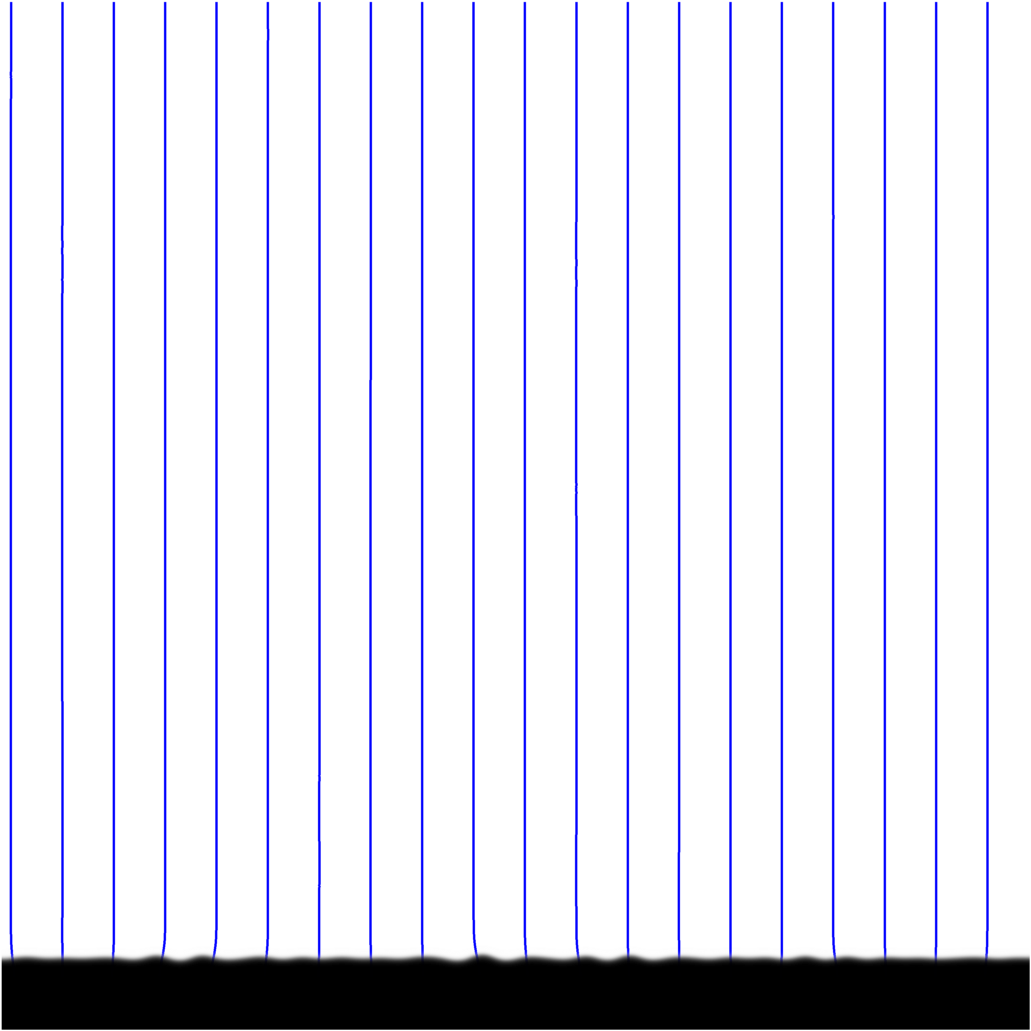} & \includegraphics[width=.18\textwidth]{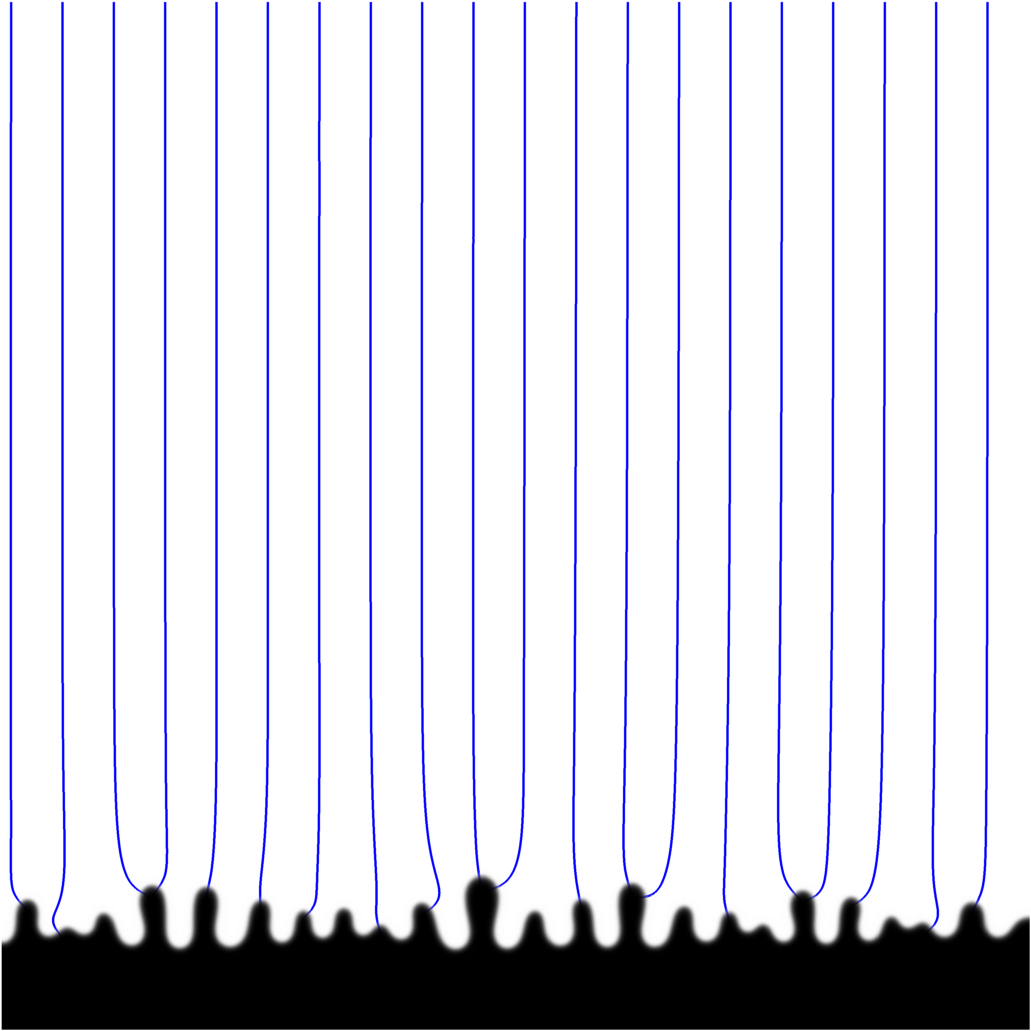} & \includegraphics[width=.18\textwidth]{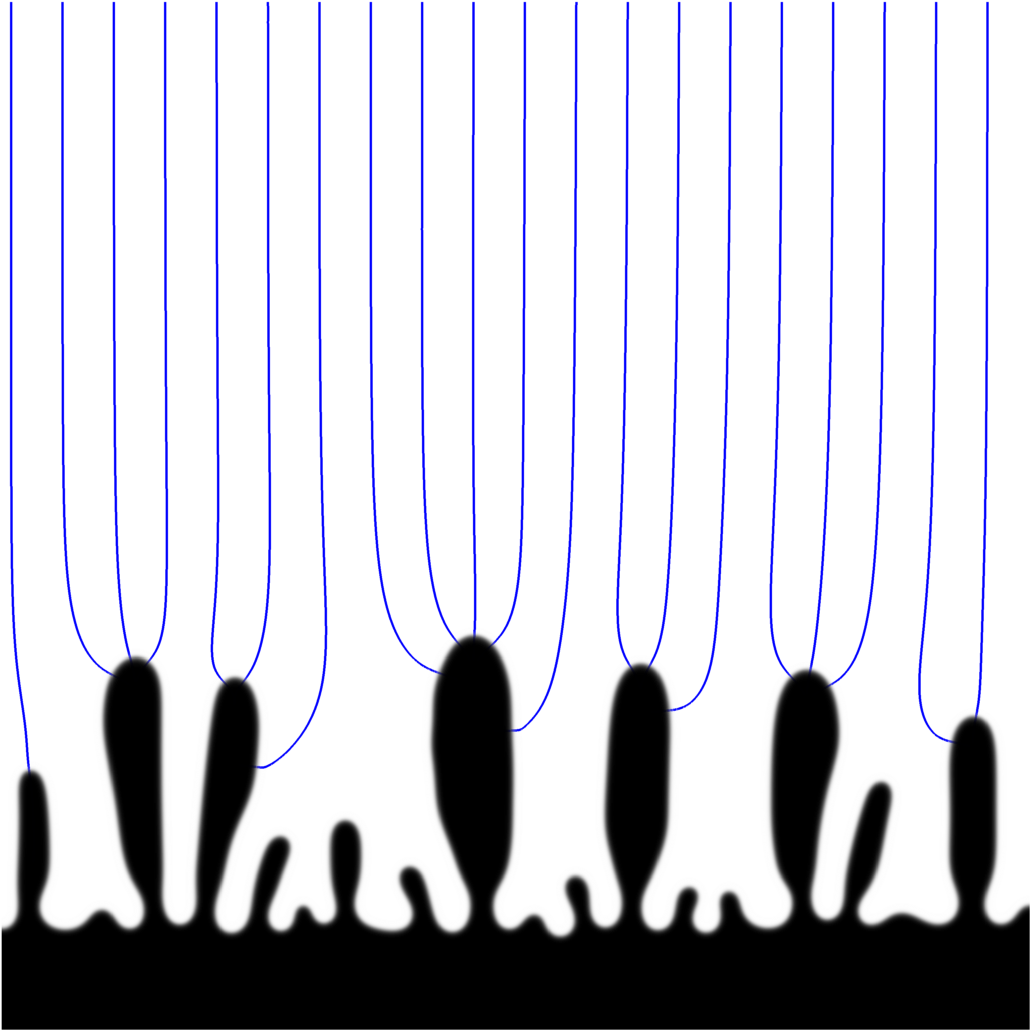} & \includegraphics[width=.18\textwidth]{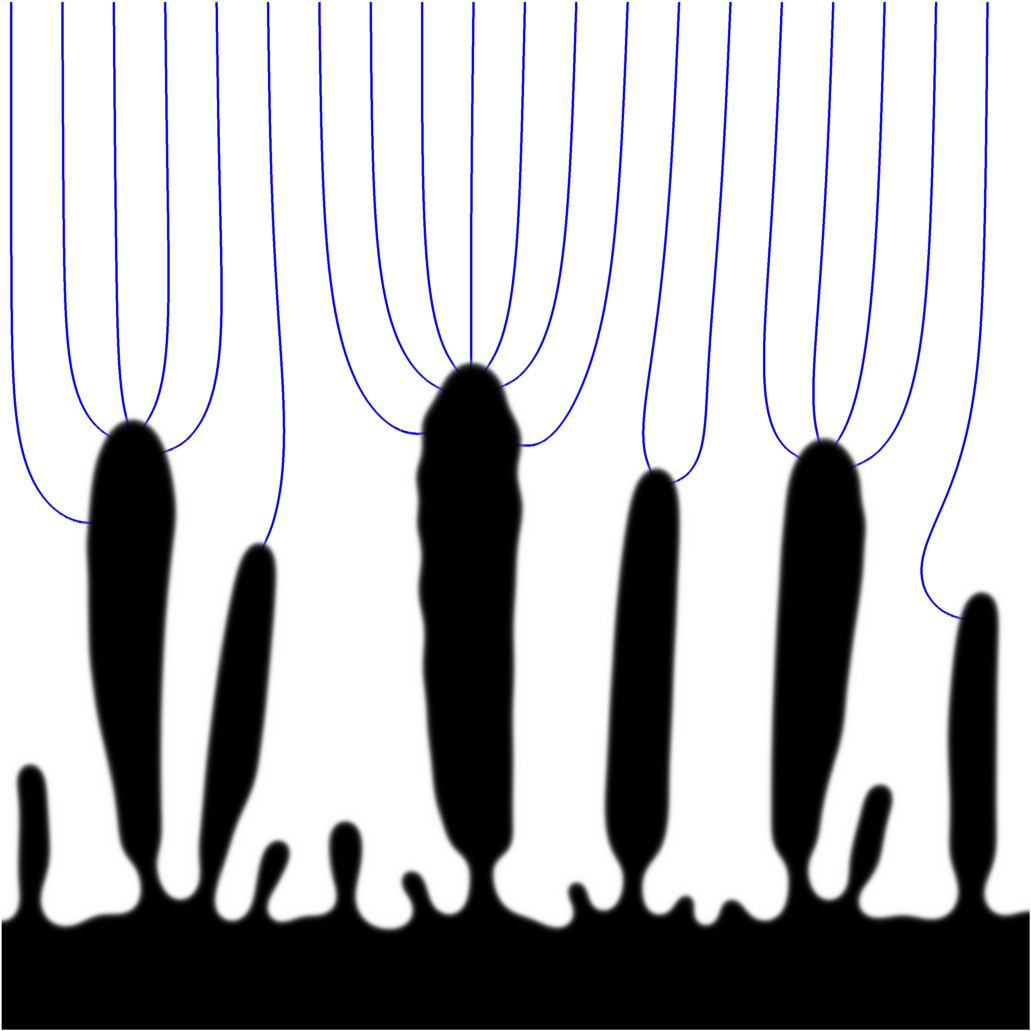}  & \includegraphics[width=.18\textwidth]{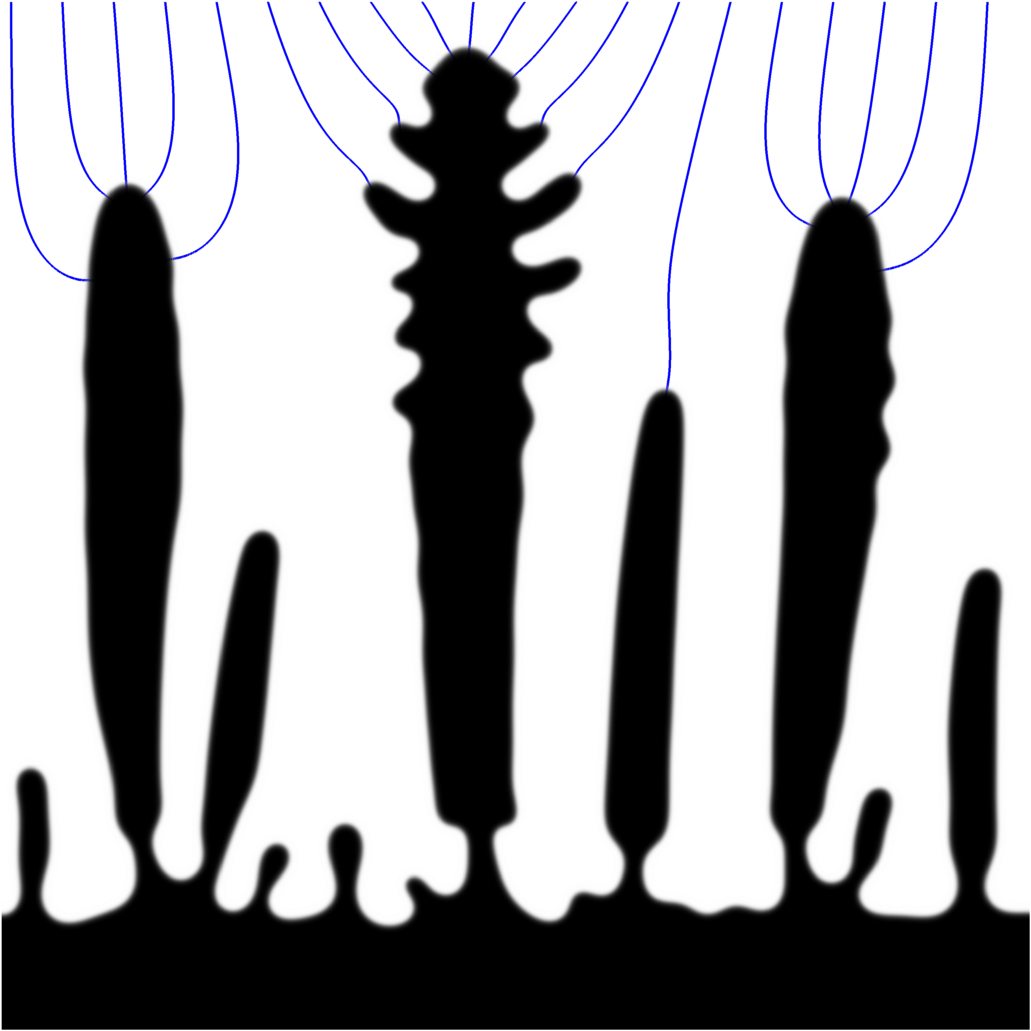} \\
 t=\unit[30]{s} & t=\unit[50]{s} & t=\unit[100]{s} & t=\unit[150]{s} & t=\unit[200]{s} \\
 \end{tabular}
 \label{Fig:cubic}}\\
 \subfloat[]{
 \begin{tabular}{ccccc}
  \includegraphics[width=.18\textwidth]{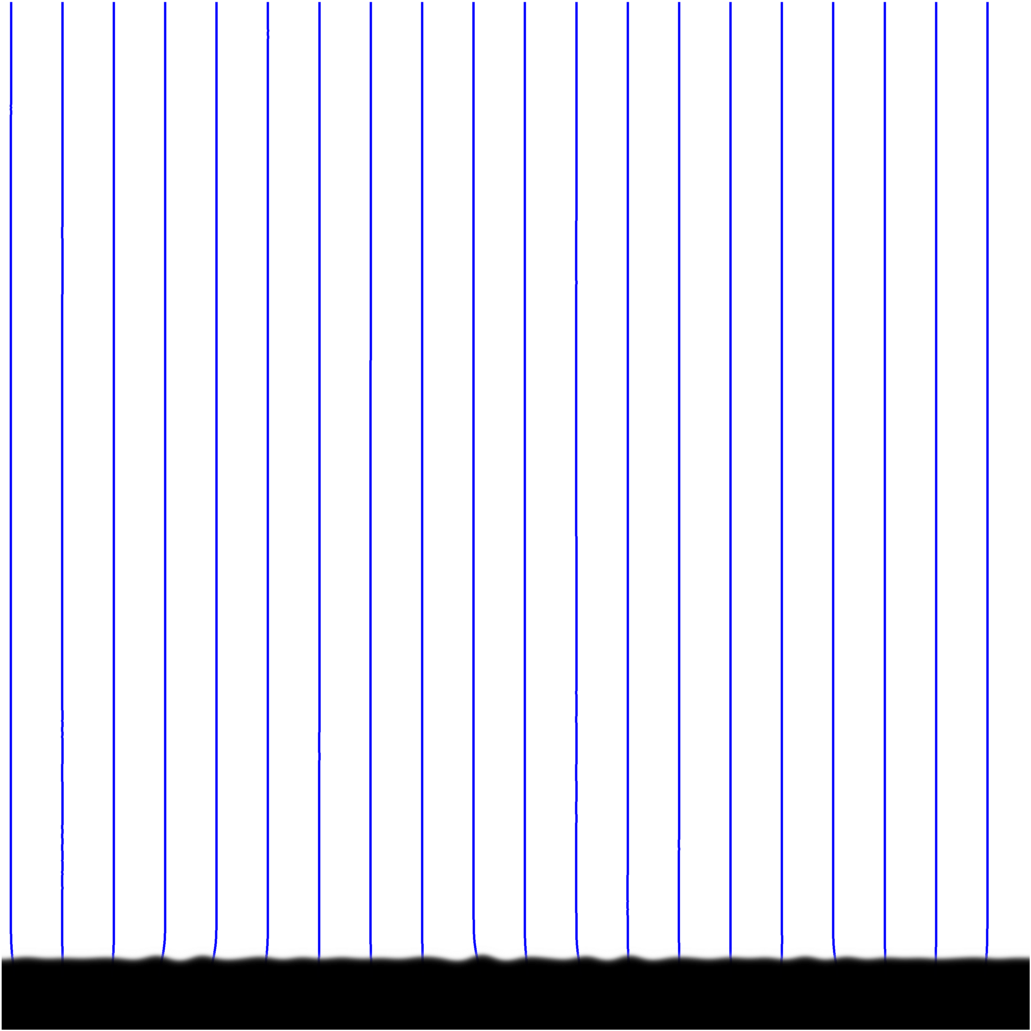} & \includegraphics[width=.18\textwidth]{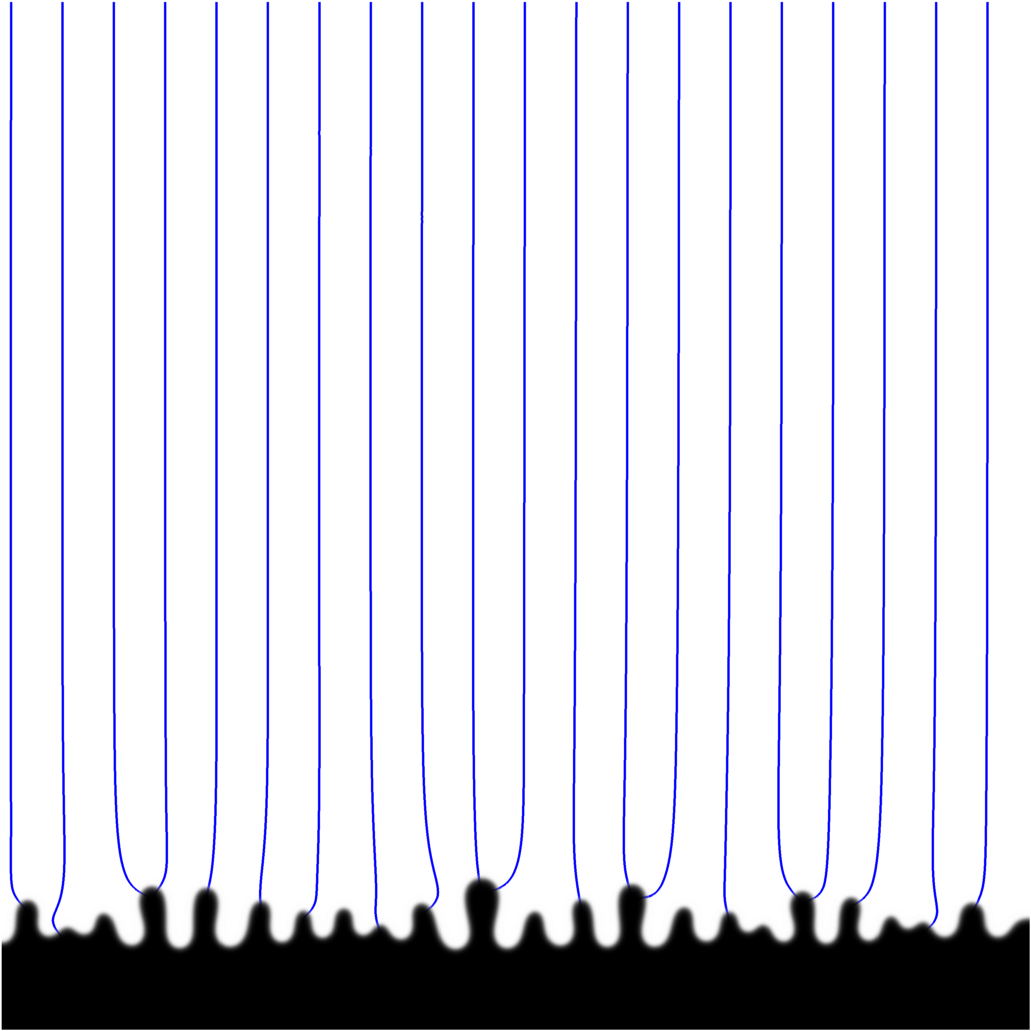} & \includegraphics[width=.18\textwidth]{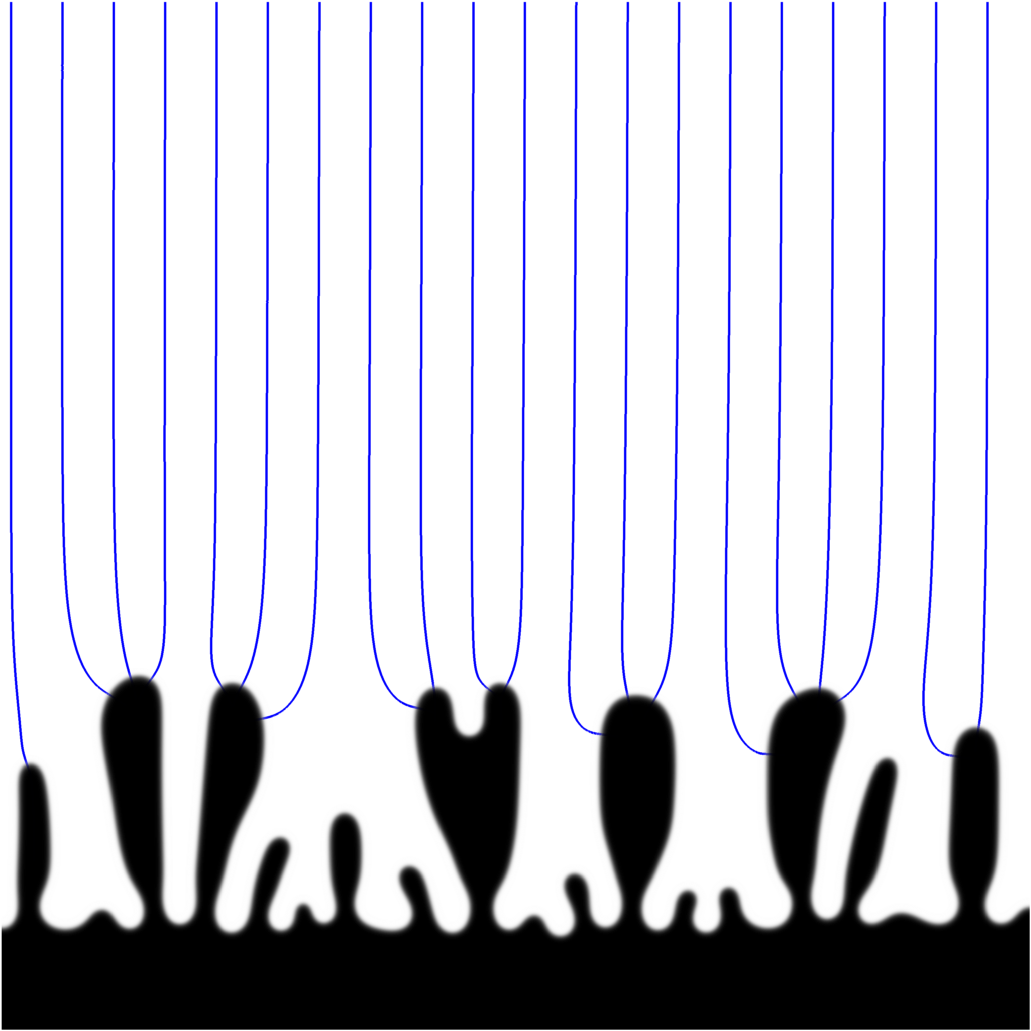} & \includegraphics[width=.18\textwidth]{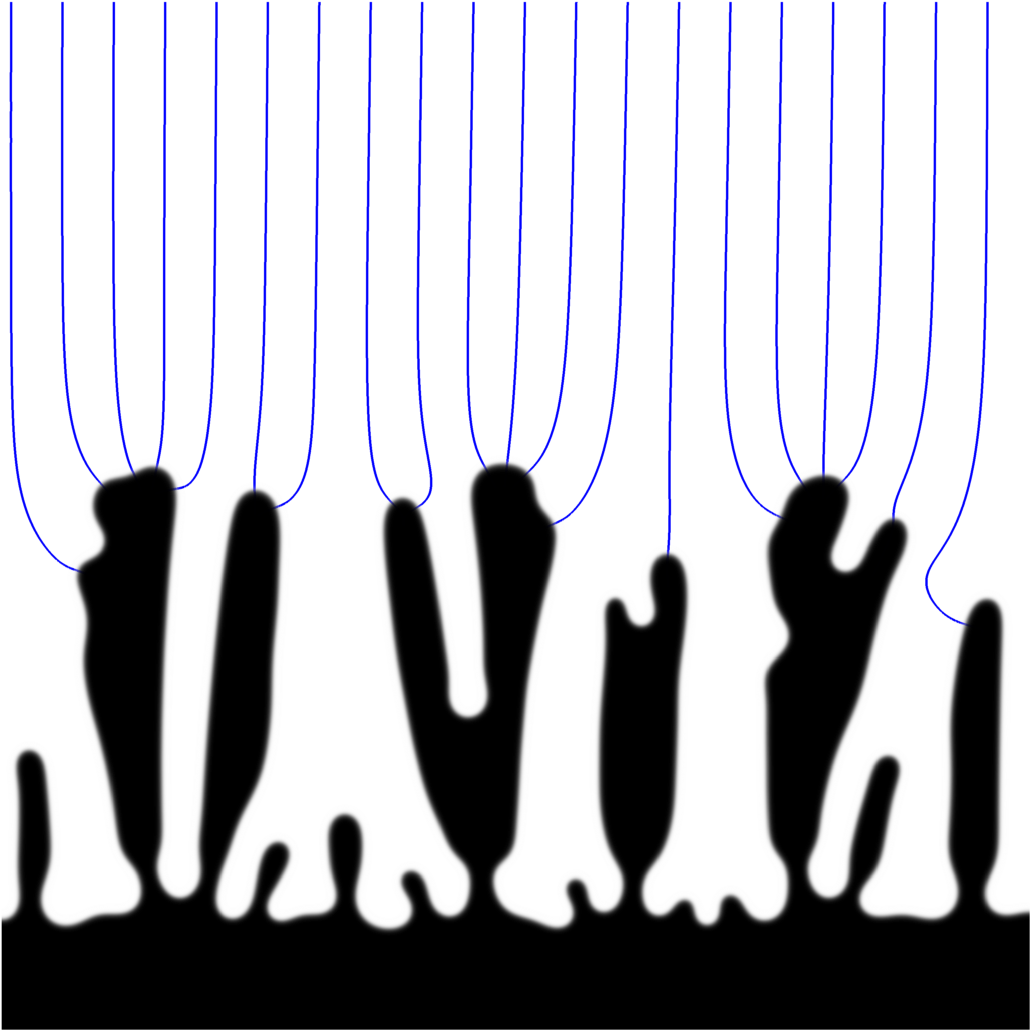} & \includegraphics[width=.18\textwidth]{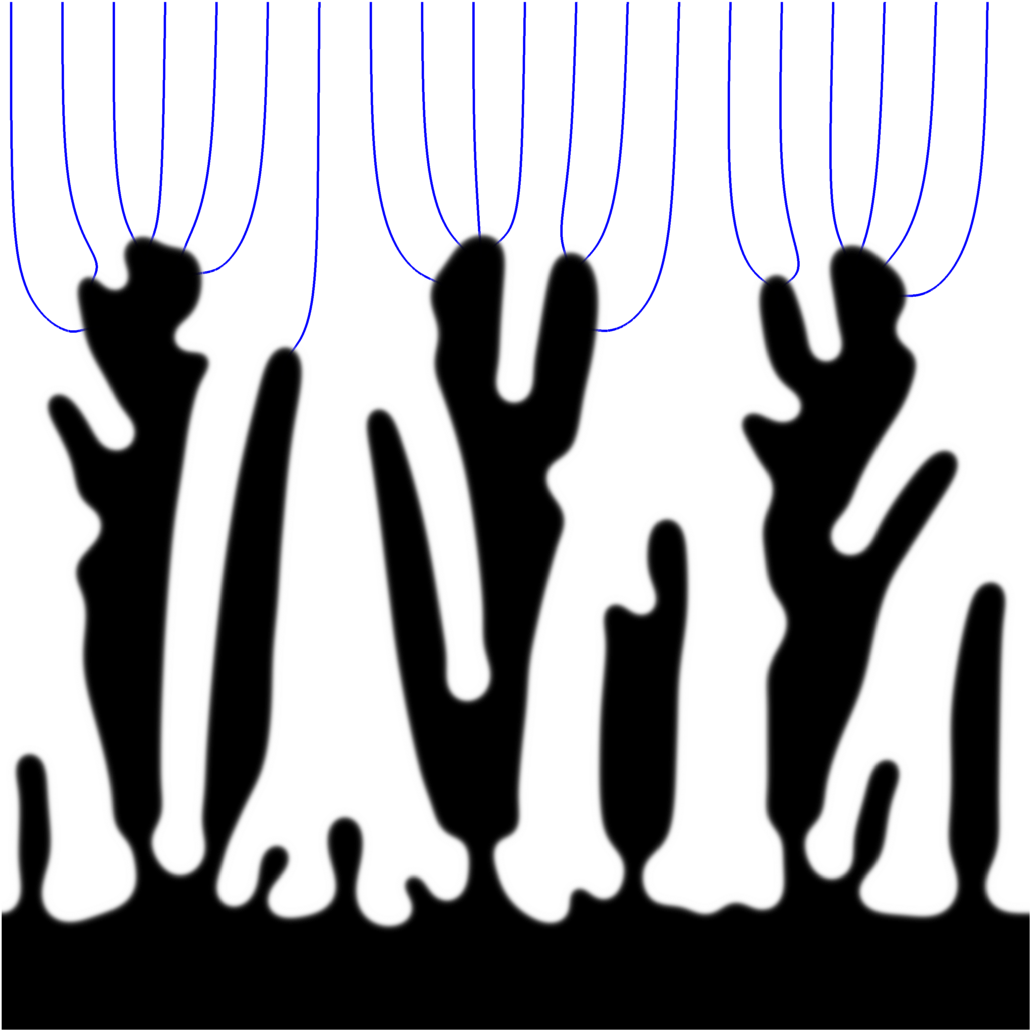} \\
 t=\unit[30]{s} & t=\unit[50]{s} & t=\unit[100]{s} & t=\unit[150]{s}  & t=\unit[200]{s} \\
 \end{tabular}
 \label{Fig:isotropic}}
\caption{Time evolution of zinc dendrite growth from a  \unit[1]{M} \ce{ZnSO4} solution with (a) hexagonal interfacial energy anisotropy ($\epsilon_6=.01$), (b) cubic anisotropy  ($\epsilon_4=.01$), and (c) isotropic surface energy. The applied electric field is \unit[1000]{V/m}, the Damkohler number is Da=10, and the width of the simulation is $\unit[150]{\mu m}$.}
\end{figure*}

Many metals, such as copper and lithium, have cubic crystal structures and are found to grow dendrites with very different morphologies than zinc. To investigate the role of crystal structure, electrodeposition simulations were additionally performed in Fig. \ref{Fig:cubic} using cubic interfacial anisotropy of the form $\gamma(\theta)=\gamma\left[1+\epsilon_4\cos(4\theta)\right]$, with $\epsilon_4=.01$. Cubic anisotropy produces straighter dendrites with less branching. As the degree of cubic anisotropy increases, branching is further reduced and the dendrites become needle-like, unlike with hexagonal symmetry.

Time evolution of dendrite growth, corresponding to Fig. 2a in the main text, is presented in Fig. \ref{Fig:hexagonal}. The initial emergence of instability bears a striking resemblance to the experiments of Elezgaray \textit{et al.} \cite{Elezgaray1998}. The electric field lines are initially uniformly distributed across the surface. As protrusions begin to develop, the field lines bend toward them, resulting in rapid growth at the tip of the protrusions. As the field at the tip continues to grow, the tip eventually splits and the current is redistributed. As this process continues, the tall dendrites are more successfully attract field lines and screen the shorter dendrites as a result. The interfaces that form tend to align along the directions of low surface energy, as defined by the interfacial anisotropy function (hexagonal for zinc).

\bibliography{electrodeposition}